\documentclass[%
 aip,
 amsmath,amssymb,
 reprint,%
]{revtex4-1}

\usepackage{graphicx}
\usepackage{dcolumn}
\usepackage{bm}

\usepackage[utf8]{inputenc}
\usepackage[T1]{fontenc}
\usepackage{mathptmx}
\usepackage{etoolbox}
\usepackage{graphicx}  
\usepackage{dcolumn}   
\usepackage{bm}        
\usepackage{verbatim}   
\usepackage{graphicx}
\usepackage{epstopdf}
\usepackage{footnote}
\usepackage{epsfig}
\usepackage{subfigure}
\usepackage{amssymb}
\usepackage{amsmath,bm}
\usepackage{xcolor}
\usepackage{float}
\usepackage{subfigure}
\usepackage[most]{tcolorbox}
\usepackage{soul}
\usepackage{color}
\usepackage[colorlinks=true,linkcolor=blue,allcolors=blue]{hyperref}%
\usepackage[framemethod=TikZ]{mdframed}
\usepackage{lipsum}
\mdfdefinestyle{MyFrame}{%
	linecolor=blue,
	outerlinewidth=2pt,
	roundcorner=20pt,
	innertopmargin=\baselineskip,
	innerbottommargin=\baselineskip,
	innerrightmargin=20pt,
	innerleftmargin=20pt,
	backgroundcolor=gray!50!white}

\makeatletter
\def\@email#1#2{%
 \endgroup
 \patchcmd{\titleblock@produce}
  {\frontmatter@RRAPformat}
  {\frontmatter@RRAPformat{\produce@RRAP{*#1\href{mailto:#2}{#2}}}\frontmatter@RRAPformat}
  {}{}
}%
\makeatother
\begin{document}

\preprint{AIP/123-QED}
 
\title{Magnet-Free Nonreciprocal Phase-Shifter Based on Time Modulation}
	\author{Sajjad Taravati and George V. Eleftheriades\\
{\small \textit{The Edward S. Rogers Sr. Department of Electrical and Computer Engineering, University of Toronto, Toronto, Ontario M5S 3H7, Canada}}\\
	Email: sajjad.taravati@utoronto.ca}

\date{\today}

\begin{abstract}
Recently, nonreciprocal phase shifters have attracted a surge of interest thanks to the advent of nonreciprocal electromagnetic systems, such as nonreciprocal metasurfaces, nonreciprocal-beam antennas, and invisibility cloaks. To overcome the limitations associated with conventional technologies for realizing nonreciprocal phase shifters and gyrators, here we propose a low-noise, lightweight, low-profile, and linear magnetless nonreciprocal phase shifter formed by two temporal loops. The proposed temporal apparatus operates based on the generation of time-harmonic signals and destructive and constructive interferences for the undesired and desired time harmonics, respectively, at different locations of the structure. An external time-harmonic modulation signal injects an effective electronic angular momentum to the system to control the phase and frequency of the two loops. Such a temporal nonreciprocal phase shifter offers low insertion loss, and a large return loss (input matching) of greater than 28.1 dB. Additionally, this nonreciprocal phase shifter possesses a reconfigurable architecture and can be directly embedded in integrated circuit (IC) technology to create high power handling and linear IC-based nonreciprocal phase shifters.
\end{abstract}

\maketitle

%

Phase shifters are key elements of wireless communication systems, biomedical systems, radars, optical wave processors, and many other applications. However, conventional passive and reciprocal phase shifters are restricted by the Lorentz reciprocity and have limited applications in the modern communication and electromagnetic systems. Recently, several exotic electromagnetic systems have been proposed by taking advantage of nonreciprocal phase shifting, including nonreciprocal metasurfaces and nonreciprocal-beam antennas. Conventionally, nonreciprocal phase shifters are created using  magneto-optical structures~\cite{okamura1985design,hung2002non,zhou2012analytical}, ferrite-based magnetic structures~\cite{adhikari2013tunable} and transistor-loaded transmission lines~\cite{ayasli1989non,ciccognani2012active,palomba2018broadband,tanaka1965active,smith1988gaas,ayasli1989field,kalialakis2000analysis,carchon2000power,taravati2021programmable,taravati2021full,karimian2020active}. However, ferrite-based magnetic structures are practically realized using bulky, costly, heavy architectures, and are not available at high frequencies. On the other hand, transistor-based nonmagnetic nonreciprocity and magneto-optical and electro-optical modulators substitute the absence of cumbersome magnetic bias for other tangible drawbacks, such as poor noise performance and strong nonlinearities in the case of transistors, or the complexity and large size in the case of optical modulators.

Recently, nonmagnetic time-modulation-based nonreciprocity has spurred a surge of research activity to eliminate the issues associated with conventional nonreciprocal devices~\cite{Taravati_Kishk_MicMag_2019,camacho2020achieving,darvish2022space,elnaggar2021properties,kumar2022computational,oudich2022bandgap}. Space-time-modulated structures take advantage of one-way progressive wave coupling to create nonreciprocal devices and isolators~\cite{Taravati_PRB_SB_2017,li2019nonreciprocal,Taravati_PRAp_2018,taravati_PRApp_2019,taravati2021lightweight,wan2021nonreciprocal}. Other applications of space-time modulation include parametric wave amplification~\cite{tien1958traveling,tien1958parametric,zhu2020tunable}, pure frequency generation~\cite{Taravati_PRB_Mixer_2018,taravati2021pure,bui2022space,gaxiola2021temporal}, nonreciprocal filters~\cite{wu2020frequency}, metasurface-based wave processors~\cite{zang2019nonreciprocal_metas,Taravati_Kishk_TAP_2019,taravati_APM_2022,khorrami2022asymmetrical,Taravati_ACSP_2022,bai2022time,fathnan2022method,zhang2022spatiotemporal,barati2022optical,ghanekar2022violation}, antennas~\cite{zang2019nonreciprocal,do2022time,zang2022magnetless,do2022time}, unidirectional beam splitting~\cite{Taravati_Kishk_PRB_2018}, and full-transceiver schemes~\cite{Taravati_AMA_PRApp_2020}.

Phase-tailored time modulation is shown to be an excellent alternative approach to realize unique functionalities and devices~\cite{zang2019nonreciprocal,taravati2020full,taravati2022lightweight,taravati2022low} in a compact fashion. Such a technique avoids the long profile that is typically required for the realization of space-time modulation~\cite{tien1958traveling,Taravati_PRB_SB_2017}. This letter proposes a nonmagnetic nonreciprocal phase shifter comprising two phase-tailored time-modulated loops. As a result, a lightweight and low-profile device can be realized thanks to the fact that only four time-modulated varactors are required. Due to the unique and optimal architecture of the proposed loop-based temporal nonreciprocal phase shifter, low insertion loss of less than 1.2~dB is achieved between its two ports, by imposing specified constructive and destructive interferences for the ports of the nonreciprocal phase shifter. Additionally, the undesired time harmonics are strongly suppressed. The nonreciprocal phase shifter exhibits highly linear response corresponding to the 1~dB compression point of +31.1~dBm, broad-band operation where a nonreciprocal phase shift with $4^\circ$ differential phase shift deviation is achieved across a 28$\%$ fractional bandwidth.

\begin{figure}
	\centering
	\subfigure{\label{fig:TL} 
		\includegraphics[width=0.995\columnwidth]{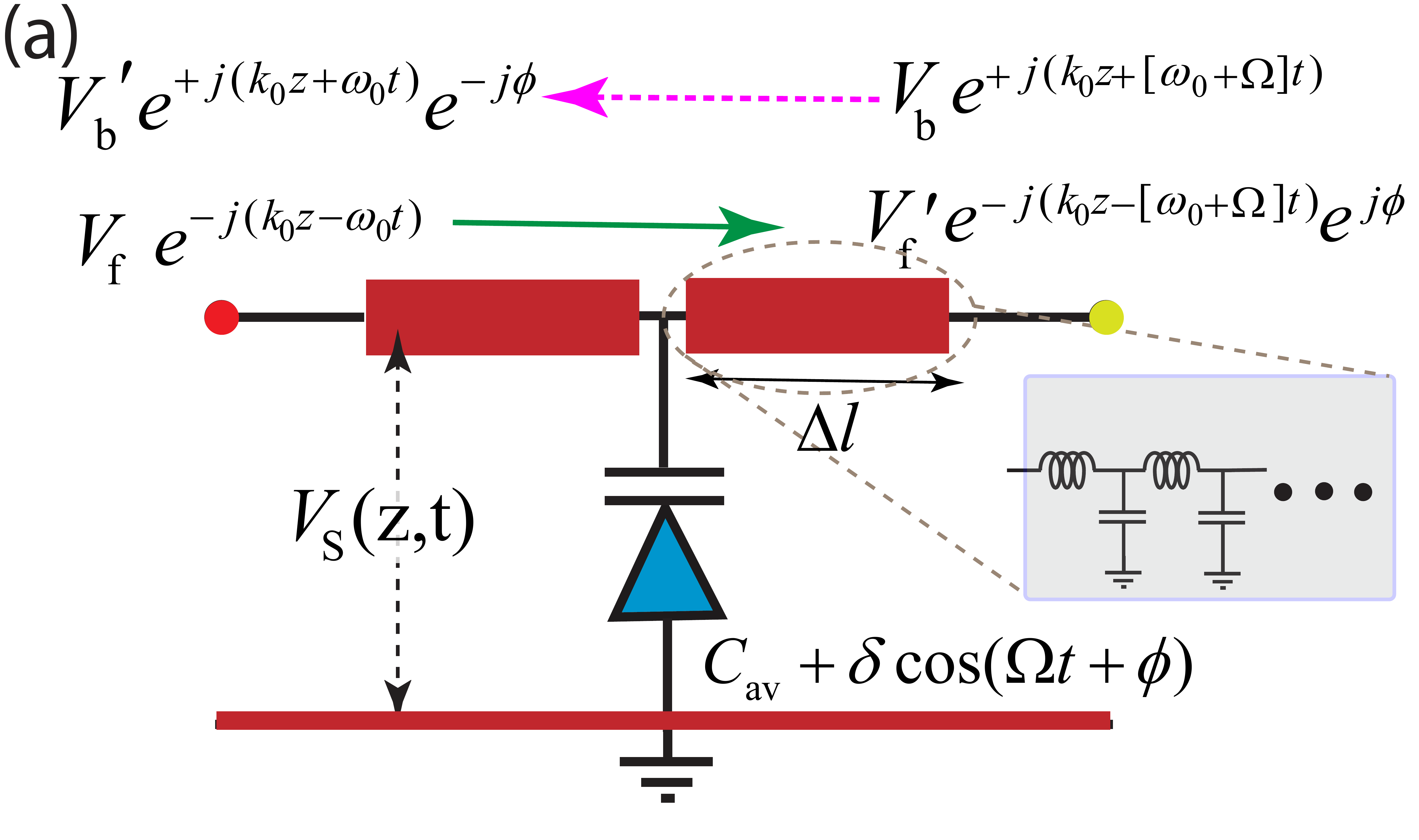}} 
	\subfigure{\label{fig:disp} 
		\includegraphics[width=0.95\columnwidth]{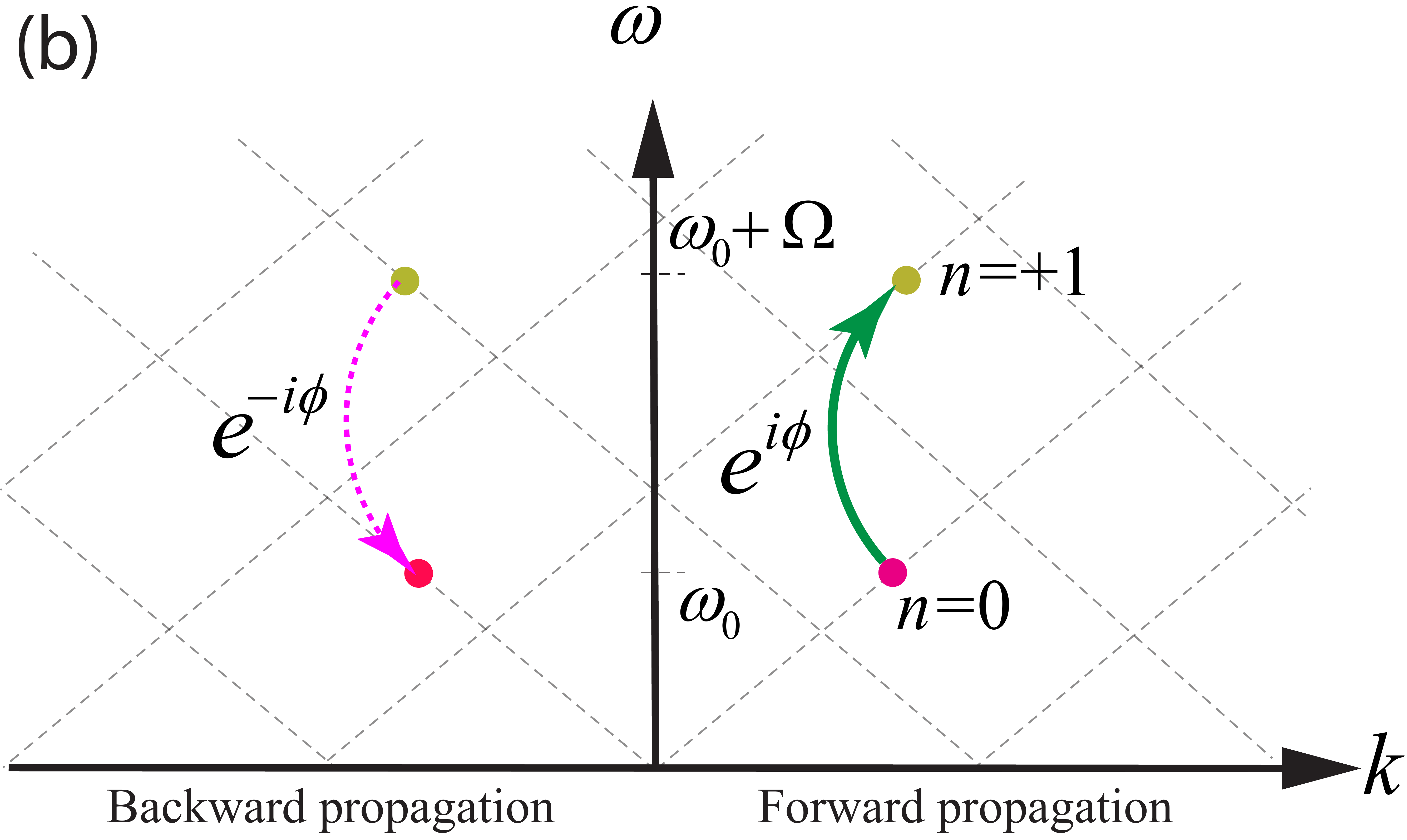}}
	\caption{Forward and backward frequency-phase transitions in a time-modulated transmission line. (a) A phased time-modulated transmission line. (b) Dispersion diagram showing forward transition from the $n=0$ time harmonic at $\omega_0$ to the $n=+1$ time harmonic at $\omega_0+\Omega$ resulting in phase alteration of $+\phi$, and backward transition from $\omega_0+\Omega$ to $\omega_0$ with phase alteration of $-\phi$.}
	\label{fig:disp1}
\end{figure}

Fig.~\ref{fig:disp1}(a) shows a transmission line loaded with a phased time-modulated varactor. Thus, the time-varying capacitance of the transmission line reads $C_\text{eq}(t)=C_\text{av} + \delta \cos(\Omega t+\phi)$, where $C_\text{av}$ represents the average capacitance of the time-varying varactor. The total capacitance of the transmission line reads $C_\text{tot}(t)=C_0C_\text{eq}(t)$, where $C_0$ is the capacitance per unit length of the transmission line. By proper engineering of the dispersion of the transmission line, the output voltage acquires a frequency transition (up/down-conversion) accompanied by a phase transition. Such frequency and phase transitions are shown in the dispersion diagram in Fig.~\ref{fig:disp1}(b), where a frequency up-conversion (from $\omega_0$ to $\omega_0+\Omega$) is associated with a phase addition of $\phi$.

The dispersion bands of the structure are tailored in a way that only the fundamental and the first higher-order time-harmonics are excited, and all undesired higher-order time-harmonics are suppressed. Hence, the space-time varying voltage between the two conductors of the transmission line is defined based on the superposition of the $n=0$ and $n=-1$ space-time harmonics fields, i.e.,
\begin{equation}\label{eqa:el}
	V_\text{S}(z,t)=a_{0}(z) e^{-i \left(k_0 z -\omega_0 t \right)}+a_{1}(z) e^{-i \left(k_0 z -(\omega_0+\Omega) t \right)}.
\end{equation}
Since the transmission line is modulated in time, and not in space, we consider an identical wave number $k_0$ for both the fundamental and the higher-order harmonics. As shown in Fig.~\ref{fig:disp1}(b), as a result of the time-modulated sinusoidally periodic capacitance, the dispersion diagram of the time-modulated transmission line is periodic with respect to the $\omega$ axis~\cite{Halevi_PRA_2009}, that is $k(\omega_0+\Omega)=k(\omega_0)=k_0$. Such a time-modulated periodicity introduces vertical electromagnetic transitions in the dispersion diagram, as demonstrated in Fig.~\ref{fig:disp1}(b). The unknown spatially variant amplitudes $a_{0}(z)$ and $a_{1}(z)$ should be determined through satisfying both the Telegrapher’s equations and the initial conditions at $z=0$. The up-conversion assumes the initial conditions of $v_{0}^{\text{up-c}}(0)=a_0$ and $v_{1}^{\text{up-c}}(0)=0$, which yields
\begin{subequations}
	\begin{equation}\label{eq:up1}
		v_{0}^{\text{up-c}}(z)=a_0 \cos\left(\dfrac{\delta k_1}{4 C_\text{av}} z\right),
\end{equation}
	\begin{equation}\label{eq:up2}
		v_{1}^{\text{up-c}}(z)=a_0\dfrac{k_1 }{k_0}  \sin\left(\dfrac{\delta k_1}{4 C_\text{av}} z\right) e^{+i \phi} .
	\end{equation}
\end{subequations}

Equations~\eqref{eq:up1} and~\eqref{eq:up2} show that the change of the frequency and phase of the up-converted signal corresponds to the frequency and phase of the time-modulated modulation signal, $\Omega$ and $+\phi$. In addition, the maximum amplitude of the up-converted signal $a_1^{\text{up-c}}(z)$ occurs at $\delta k_1/ C_\text{av}=2\pi$. This reveals that, one can control the amplitude of the up-converted signal by changing the modulation amplitude $\delta$. Following the same procedure, we can determine the input and output voltages for the down-conversion. 

The down-conversion assumes the initial conditions of $v_{0}(0)=0$ and $v_{1}^{\text{dn-c}}(0)=a_1=\text{Max}[v_{1}^{\text{dn-c}}(z)]=a_0 k_1 /k_0$, which gives
\begin{subequations}
	\begin{equation}\label{eq:down1}
		v_{1}^{\text{dn-c}}(z)=a_1 \cos\left(\dfrac{\delta k_1}{4 C_\text{av}} z\right),
	\end{equation}
	and
	\begin{equation}\label{eq:down2}
		v_{0}^{\text{dn-c}}(z)= a_0    \sin\left(\dfrac{\delta k_1}{4 C_\text{av}} z \right) e^{-i \phi}.
	\end{equation}
\end{subequations}

Equations~\eqref{eq:down1} and~\eqref{eq:down2} show that the change of the frequency and phase of the down-converted signal corresponds to the frequency and phase of the time-modulated modulation signal, $\Omega$ and $-\phi$.

\begin{figure}
	\centering
	\includegraphics[width=0.98\columnwidth]{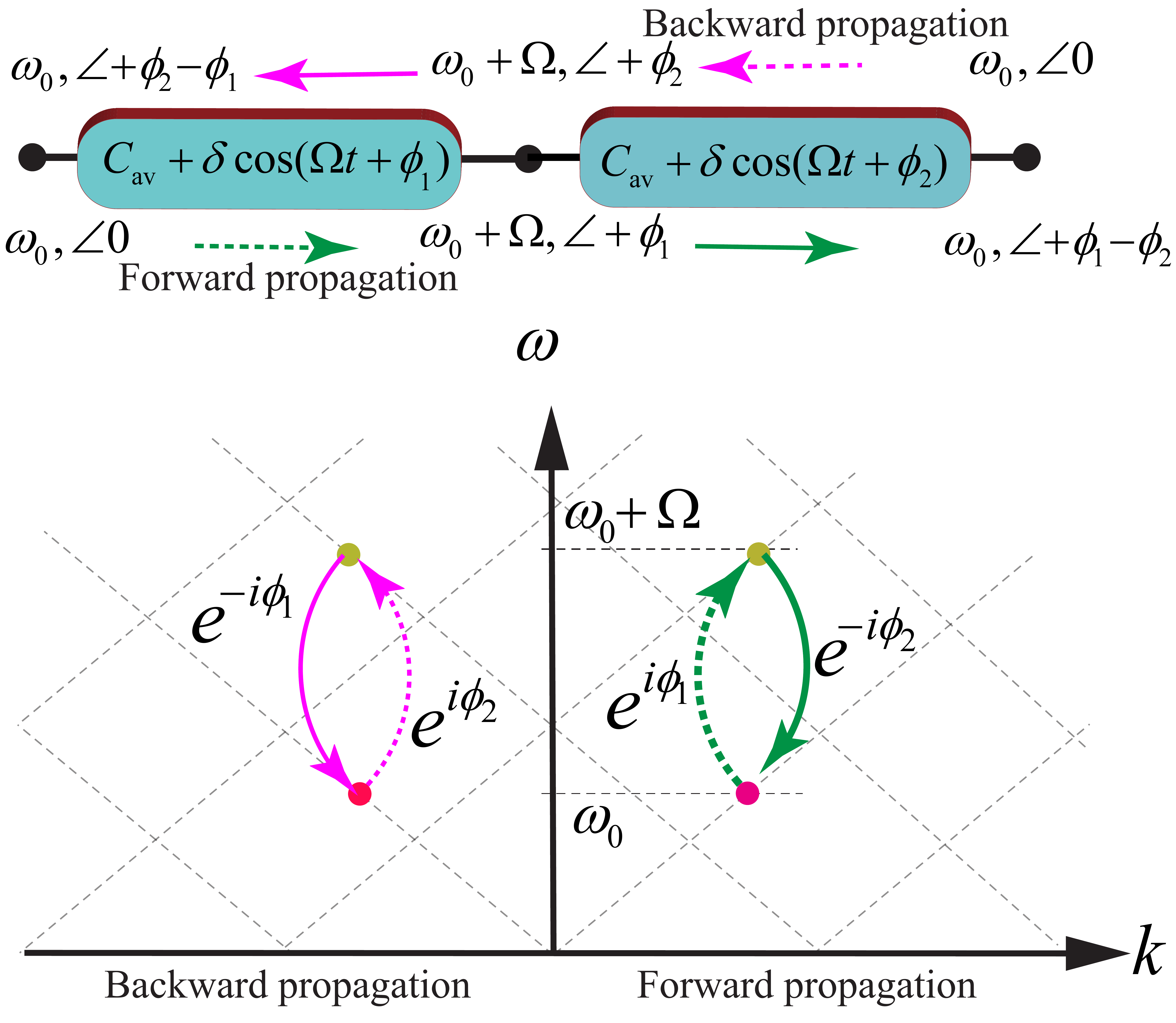}
	\caption{Schematic representation of forward and backward frequency-phase transitions in two cascaded time-modulated transmission lines.}
	\label{fig:disp2}
\end{figure}

Figure~\ref{fig:disp2} shows a schematic representation of the dispersion diagram for asymmetric frequency-phase transitions in a time-modulated transmission line. Here, a frequency up-conversion from $\omega_0$ to $\omega_0+\Omega$ is accompanied by a positive additive phase, i.e., $+\phi_1$ in the forward transmission and $+\phi_2$ in the backward transmission. However, a frequency down-conversion from $\omega_0+\Omega$ to $\omega_0$ is accompanied by a negative additive phase, i.e., $-\phi_2$ in the forward transmission and $-\phi_1$ in the backward transmission.

\begin{figure*}
	\centering
	\includegraphics[width=1.5\columnwidth]{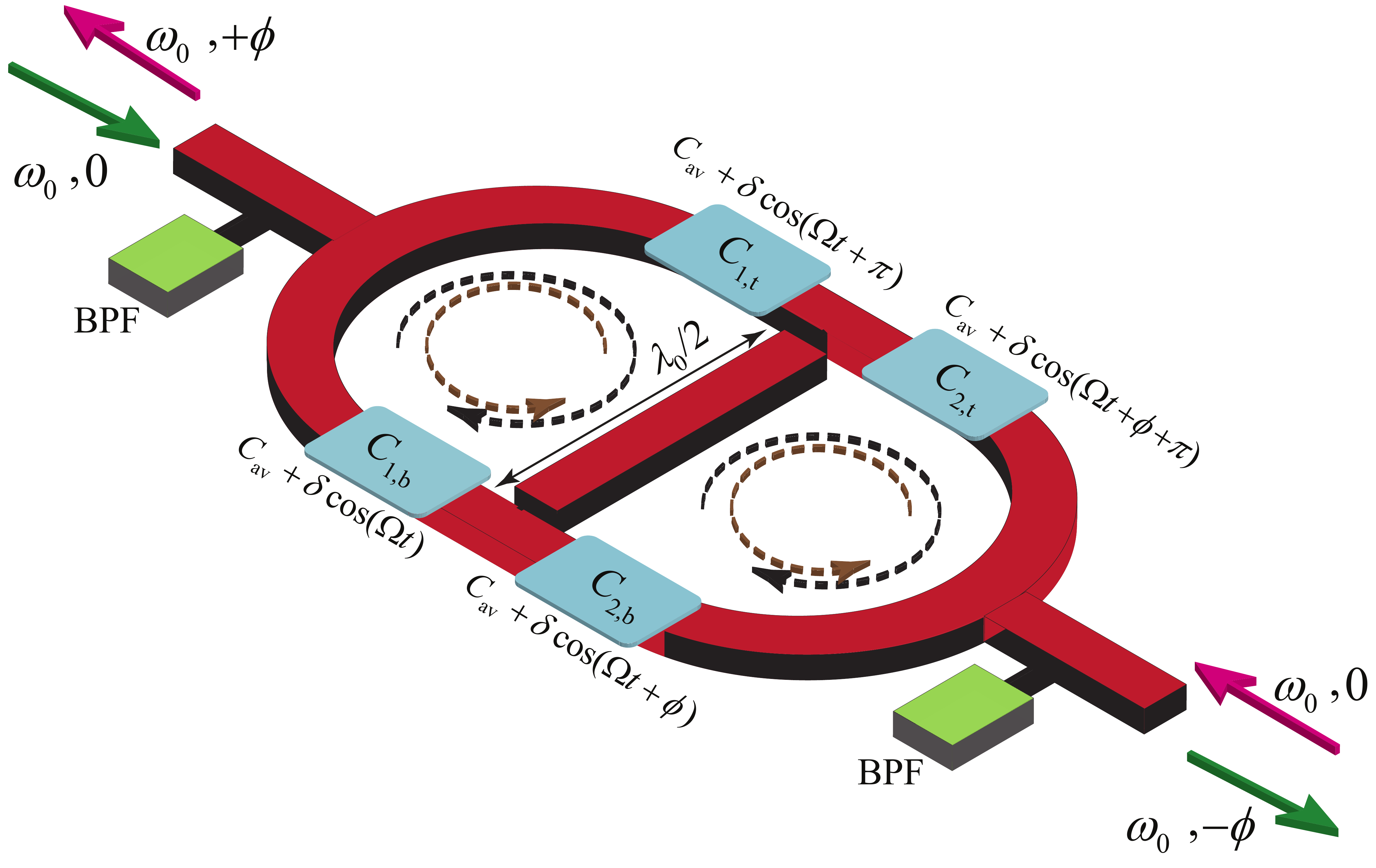}
	\caption{Architecture of the nonreciprocal phase shifter formed by two phase-tailored time-modulated loops.}
	\label{fig:sch}
\end{figure*}

The time-varying nonreciprocal phase shifting architecture in Fig.~\ref{fig:disp2} assumes only two time harmonics are supported by the transmission line, while all the undesired time harmonics are not excited. In reality, however, this is not true, and all the time harmonic side-bands are generated which degrades the performance of nonreciprocal phase shifting. To overcome this issue, here we propose a new architecture that is constituted of four temporal unit cells in the form of two loops. Such an architecture guarantees propagation and transmission of desired time harmonics and eliminates the undesired time harmonic side-bands, leading to an excellent nonreciprocal phase shifting performance. Figure~\ref{fig:sch} shows the proposed phase-engineered loop-based time-modulated nonreciprocal phase shifter, formed by two time-varying loops, each of which is composed of two phase-engineered time-modulated transmission lines. In comparison with the simple nonreciprocal phase shifting platform in Fig.~\ref{fig:disp2}, the loop-based architecture in Fig.~\ref{fig:sch} introduces three junctions between its top and bottom arms that are utilized for governing the time harmonics, that is, forming constructive interferences at the desired time harmonics and destructive interferences at the undesired time harmonics, as will be explained later. 

The left time-modulated loop is composed of two time-modulated transmission lines, where the capacitance of the transmission lines is modulated in time introducing a $\pi$ phase difference between the bottom and top time-modulated cells. The right time-modulated loop is composed of two time-modulated transmission lines, where the capacitance of the transmission lines is modulated in time introducing a $\pi$ phase difference between the bottom and top time-modulated cells and $\phi$ phase difference with the left loop, that is,	
\begin{subequations}\label{fig:capac}
	\begin{equation}\label{fig:capac_a}
		C_{1,\text{b}}(t)=C_\text{av}+\delta \cos(\Omega t),
	\end{equation}
	\begin{equation}\label{fig:capac_b}
		C_{1,\text{t}}(t)=C_\text{av}+\delta \cos(\Omega t+\pi),
	\end{equation}
	\begin{equation}
		C_{2,\text{b}}(t)=C_\text{av}+\delta \cos(\Omega t+\phi),
	\end{equation}
	\begin{equation}
		C_{2,\text{t}}(t)=C_\text{av}+\delta \cos(\Omega t+\phi+\pi).
	\end{equation}
\end{subequations}

In Eq.~\eqref{fig:capac}, $C_\text{av}$ represents the average capacitance of the transmission line loaded with time-modulated capacitance, $\delta$ is the strength of the time modulation, $\Omega$ is the modulation frequency, and $\phi$ represents the modulation phase. Next, by proper design of the time modulation of the two loops, a desired nonreciprocal phase shift is achieved. These two time-modulated loops introduce constructive interferences for the desired time harmonics and destructive interferences for the undesired time harmonics. Two band-pass filters (BPF) at $\omega_0$ are used at the two ports of the nonreciprocal phase shifter to further suppress and eliminate the undesired side-band time harmonics. 

\begin{figure*}
	\begin{center}
		\subfigure{\label{fig:p1} 
			\includegraphics[width=1.3\columnwidth]{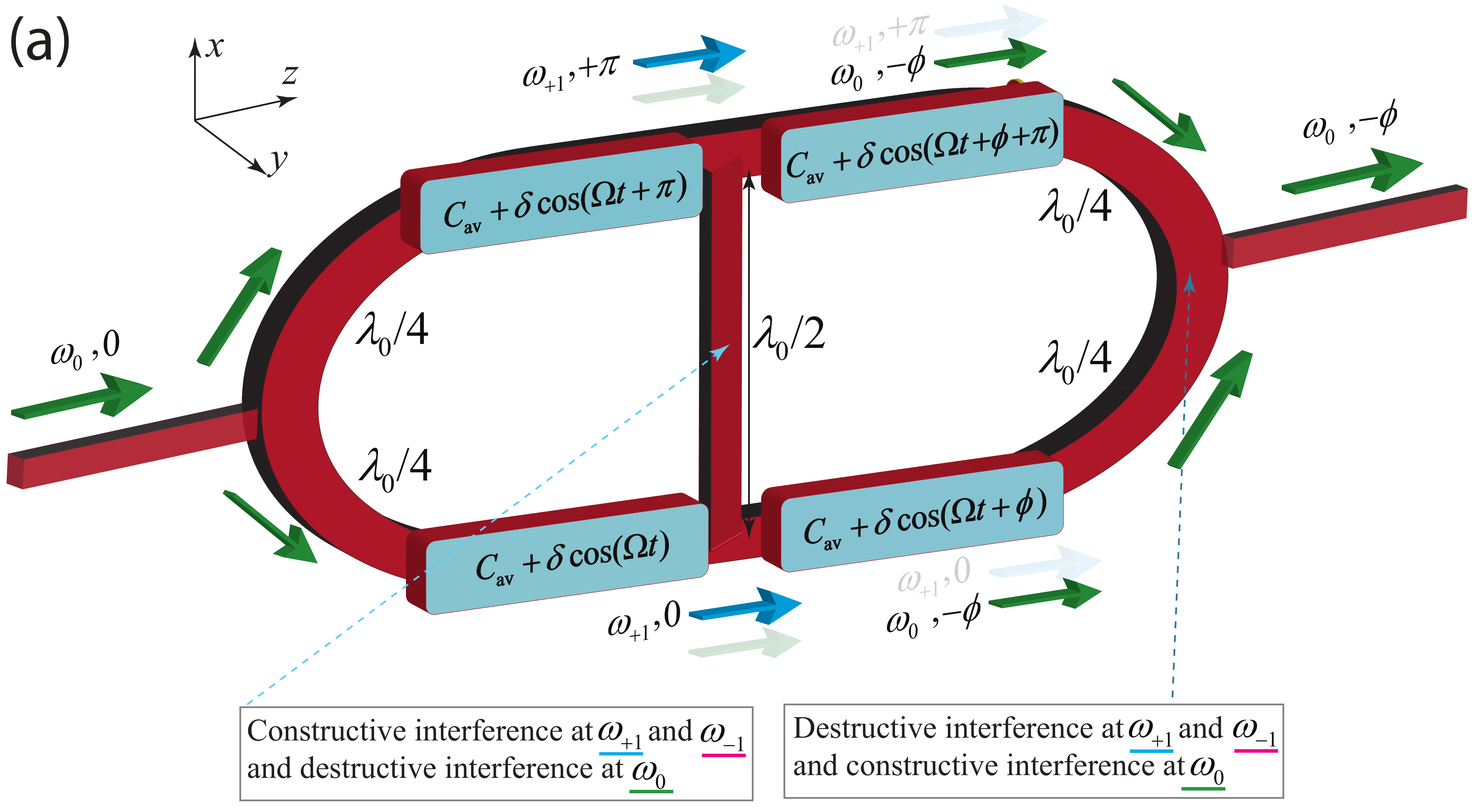}} 
		\subfigure{\label{fig:p2} 
			\includegraphics[width=1.3\columnwidth]{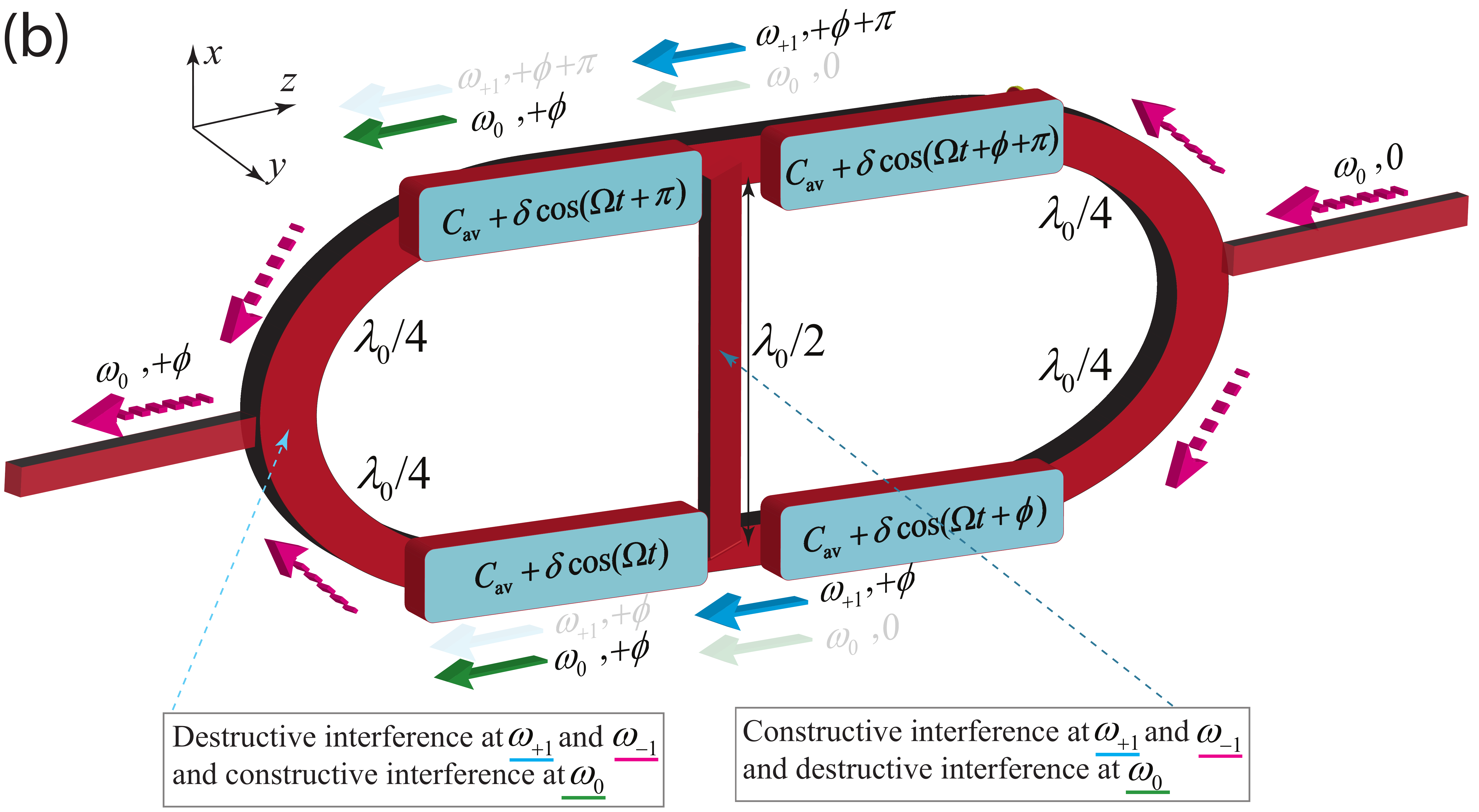}}
		\caption{Appropriate constructive and destructive signal interferences in the two loops are achieved through engineered frequency and phase transitions in the proposed time-modulated nonreciprocal phase shifter. (a) Forward signal transmission introducing a phase alteration of $-\phi$ at the same frequency $\omega_0$. (b) Backward signal transmission introducing a phase alteration of $+\phi$ at the same frequency $\omega_0$.} 
		\label{Fig:pp}
	\end{center}
\end{figure*}

Figures~\ref{fig:p1} and~\ref{fig:p2} illustrate the signal transmission through the two arms of the nonreciprocal phase shifter under forward and backward signal injections, respectively. Two power splitters/combiners possessing two quarter-wavelength arms are placed at the two sides of the structure to govern the desired constructive and destructive interferences. In addition, a half-wavelength transmission line is placed at the middle of the structure to separate the two loops from each other and provide desired constructive and destructive interferences for different time harmonics. We aim to achieve a two-way complete signal transmission, but with opposite phase shifts for the forward and backward transmissions. This is accomplished by proper engineering of the forward (Figures~\ref{fig:p1}) and backward (~\ref{fig:p2}) frequency and phase transitions in the proposed time-modulated nonreciprocal phase shifter. 

Fig.~\ref{fig:p1} shows the forward signal injection and transmission. Here, the input signal is fed to the common port of the left power splitter and propagates through the two arms of the left loop. These two signals make a transition to the first higher order time harmonic $\omega_{+1}$ but with different phases, that is, $+\pi$ and $0$ phase shifts in the upper and lower arms, respectively. The half-wavelength interconnector at the middle of the structure makes a constructive interference at $\omega_{+1}$, and therefore, the generated $\omega_{+1}$ time harmonic excites the right side loop. This leads to the generation of the $\omega_0$ harmonic by the right side loop, possessing identical phase shift of $-\phi$ in the upper and lower arms of the output of the right side loop. As a consequence, since the power splitter is formed by equal quarter-wavelength long arms, the signals add up at the common port of the right side power divider. This results in a transmission at $\omega_0$ with the phase shift of $-\phi$. The higher order time harmonics $\omega_{+1}=\omega_{0}+\Omega$ and $\omega_{-1}=\omega_{0}-\Omega$ acquire a $\pi$ phase shift difference at the two arms of the output (right) power divider, and hence, they form a complete null at the output of the structure, and therefore are not be transmitted.

Figure~\ref{fig:p2} illustrates the backward signal transmission, where the input signal is injected to the common port of the right side power divider and propagates through the two arms of the right loop. Here, the input signal is fed to the common port of
the right power splitter and splits into two signals at two output ports of the power splitter. Next, these two signals transition to the higher order time harmonic $\omega_{+1}$ but with $+\phi+\pi$ and $+\phi$ phase shifts in the upper and lower arms, respectively. Similar to the forward signal transmission case, the half-wavelength transmission in the middle interconnector makes a constructive interference at $\omega_{+1}$ and a destructive interference at $\omega_{0}$. Hence, the generated $\omega_{+1}$ time harmonic reaches to the left side loop. This leads to the generation of the $\omega_0$ harmonic by the left side loop, with identical phase shift of $+\phi$ in the upper and lower arms of the output of the right side loop. Hence, since the power divider is formed by equal quarter-wavelength long arms, the signals add up at the common port of the left side power combiner. This results in a transmission at $\omega_0$ with the phase shift of $+\phi$ which is the opposite of the phase shift of the forward direction. Since the higher order time harmonics $\omega_{+1}$ and $\omega_{-1}$ acquire a $\pi$ phase shift difference at the two arms of the output power divider, they form a complete null at the output of the left power combiner. The difference between the phase shift of the forward and backward transmissions arises due to the asymmetric frequency-phase transitions in the proposed phase-tailored time-modulated cells.

\begin{figure*}
	\centering
	\includegraphics[width=1.3\columnwidth]{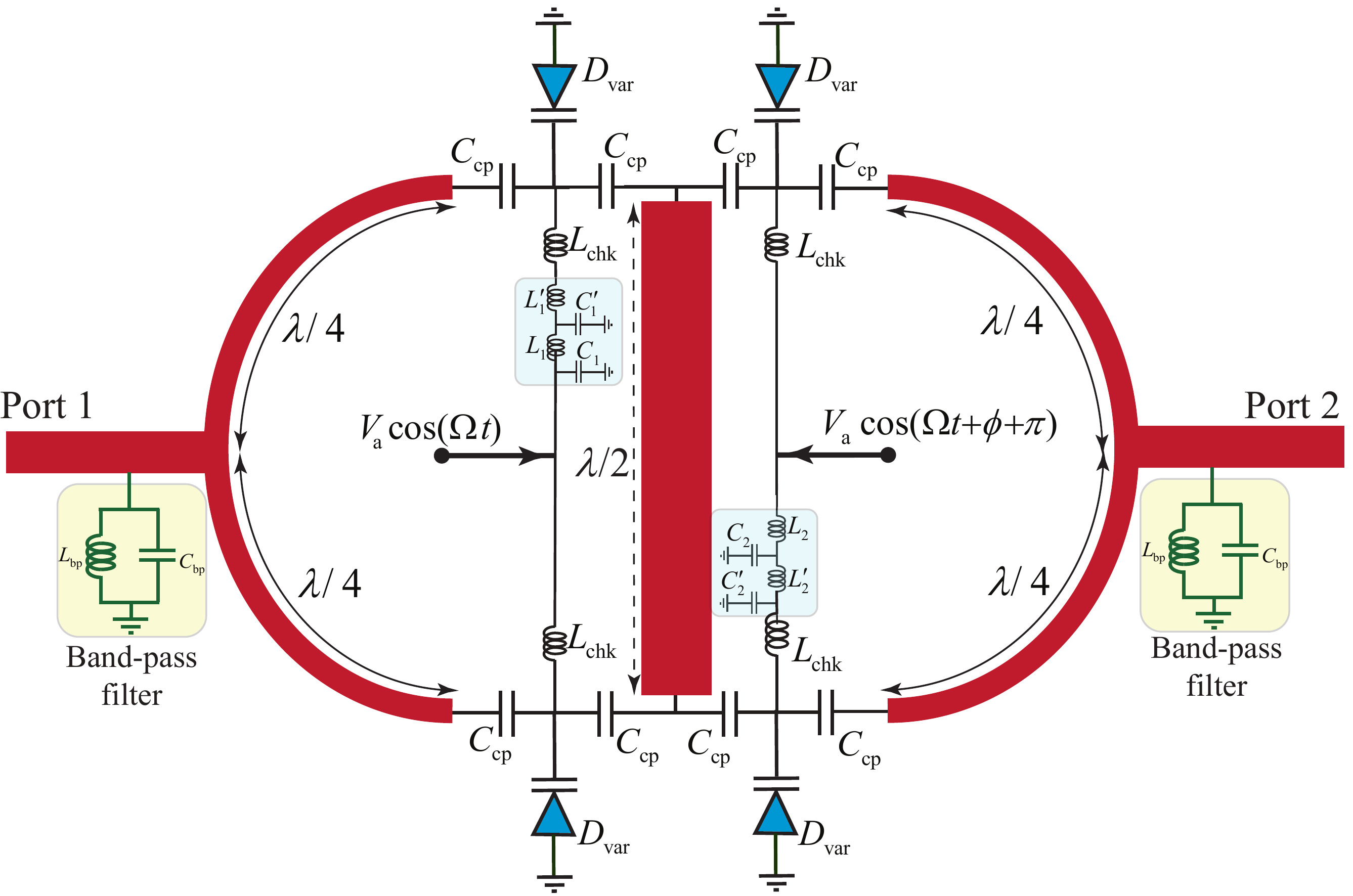}
	\caption{Practical realization of the time-modulated nonreciprocal phase shifter by four time-modulated varactors.}
	\label{fig:2}
\end{figure*}

Figure~\ref{fig:2} shows a proof of concept fabricated time-modulated nonreciprocal phase shifter using four phase-shifted time-varying varactors. Here, the two quarter-wavelength transmission lines on top of the figure form the two fixed reciprocal $90^\circ$ phase shifters.  The nonreciprocal time-modulated phase shifter is constituted of the following components. Four varactors are used to create a phase-shifted time modulation. Two transmission-line-based power splitters feed the four time-modulated transmission lines, and the half-wavelength transmission line in the middle of the nonreciprocal time-modulated phase shifter provides constructive interference for the desired time harmonics and destructive interference for the undesired time harmonics. Furthermore, four fixed inductors denoted by $L_\text{chk}=390$ nH prevent leakage of the main microwave signal to the modulation line and phase shifters, eight fixed capacitances denoted by $C_\text{cp}=10$ pF decouple the dc bias and low frequency modulation signals from the incident and transmitted high frequency incident microwave signals.


\begin{figure}
	\centering
	\subfigure{\label{fig:ph_front} 
		\includegraphics[width=80mm]{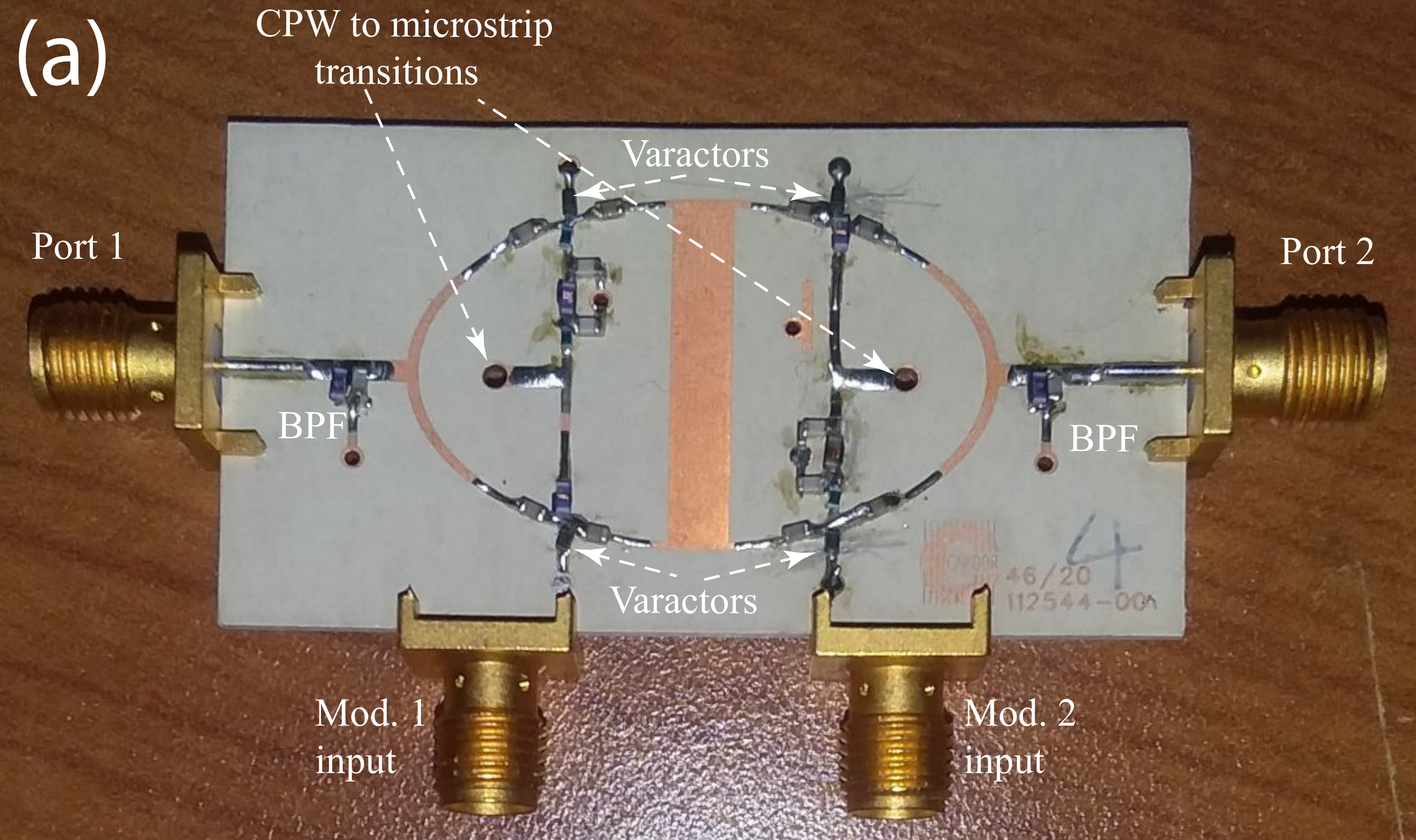}}
	\subfigure{\label{fig:ph_back}  
		\includegraphics[width=80mm]{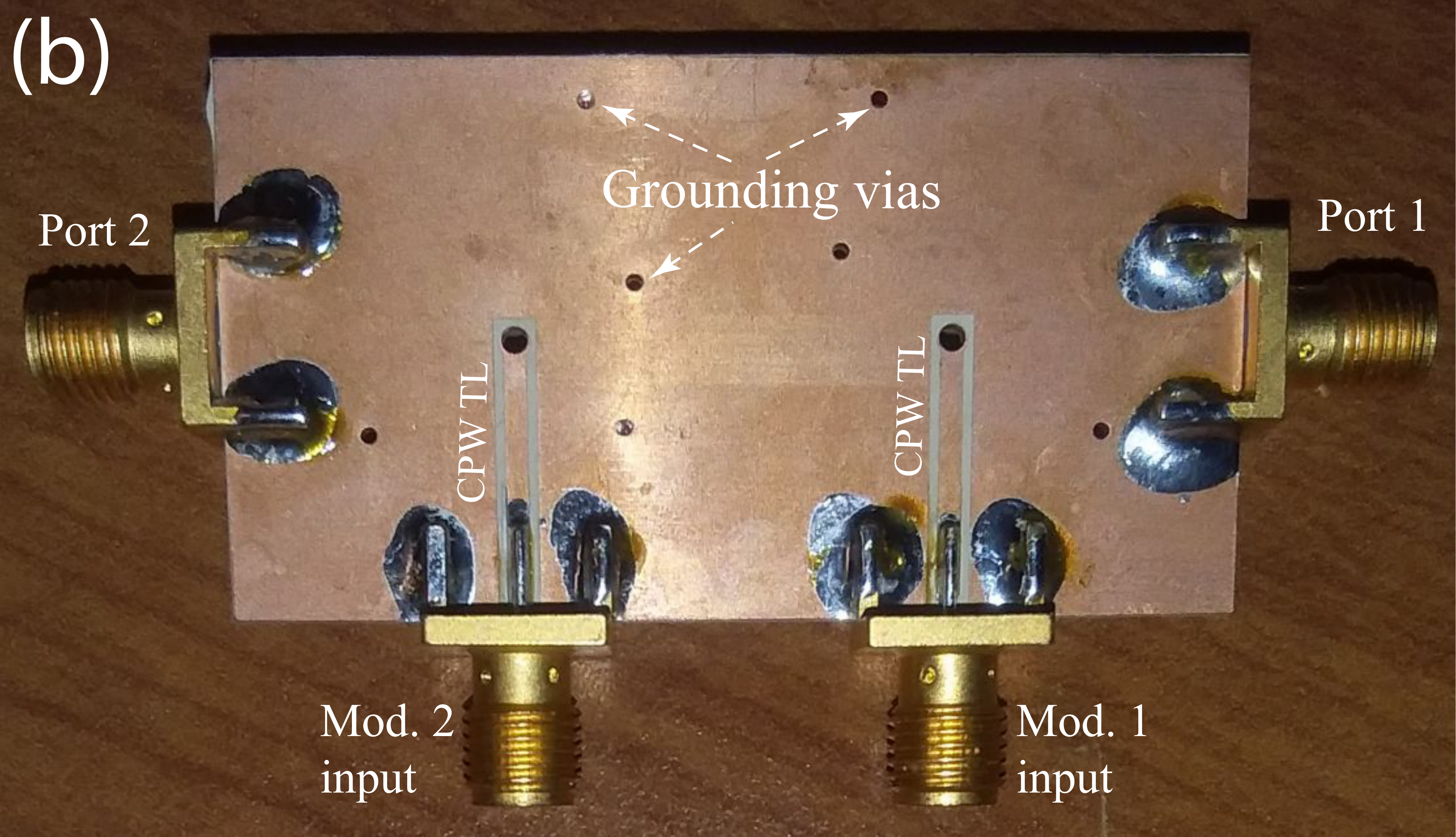}}	
	\caption{Proof of concept fabricated prototype at microwave frequencies. (a) Top view. (b) Bottom view.}
	\label{fig:3}
\end{figure}

Figures~\ref{fig:ph_front} and~\ref{fig:ph_back} show the top and bottom views of the fabricated nonreciprocal phase shifter, respectively. The nonreciprocal phase shifter is designed at the center frequency of 3.5 GHz with the modulation frequency of $\Omega=220$ MHz and modulation signal power of +15 dBm. We have utilized a RT6010 substrate with permittivity $\epsilon_\text{r} = 10.2$, thickness $h = 1.27$~mm and $\tan \delta = 0.0023$. In addition, four SMV2019-079LF Skyworks Solutions Inc. varactors are placed on top of the substrate and are grounded through four via holes. Ports 1 and 2 are the main high frequency ports of the nonreciprocal phase shifter (support 3.5 GHz), while two other ports at the bottom of the structure, that is, Mod. 1 input and Mod. 2 input, provide the low frequency modulation signal injection with different phase shifts. As shown in Fig.~\ref{fig:ph_back}, the two modulation signals, with same amplitudes and frequencies but different phases, are injected to the structure through the two 50 Ohm coplanar waveguide (CPW) transmission lines at the back of the structure. Two ZFBT-6GQ+ minicircuits bias-tees are used to concurrently feed the modulation signals and DC bias to the varactors.

\begin{figure}
	\begin{center}			
		\subfigure{\label{fig:measf} 
			\includegraphics[width=0.9\columnwidth]{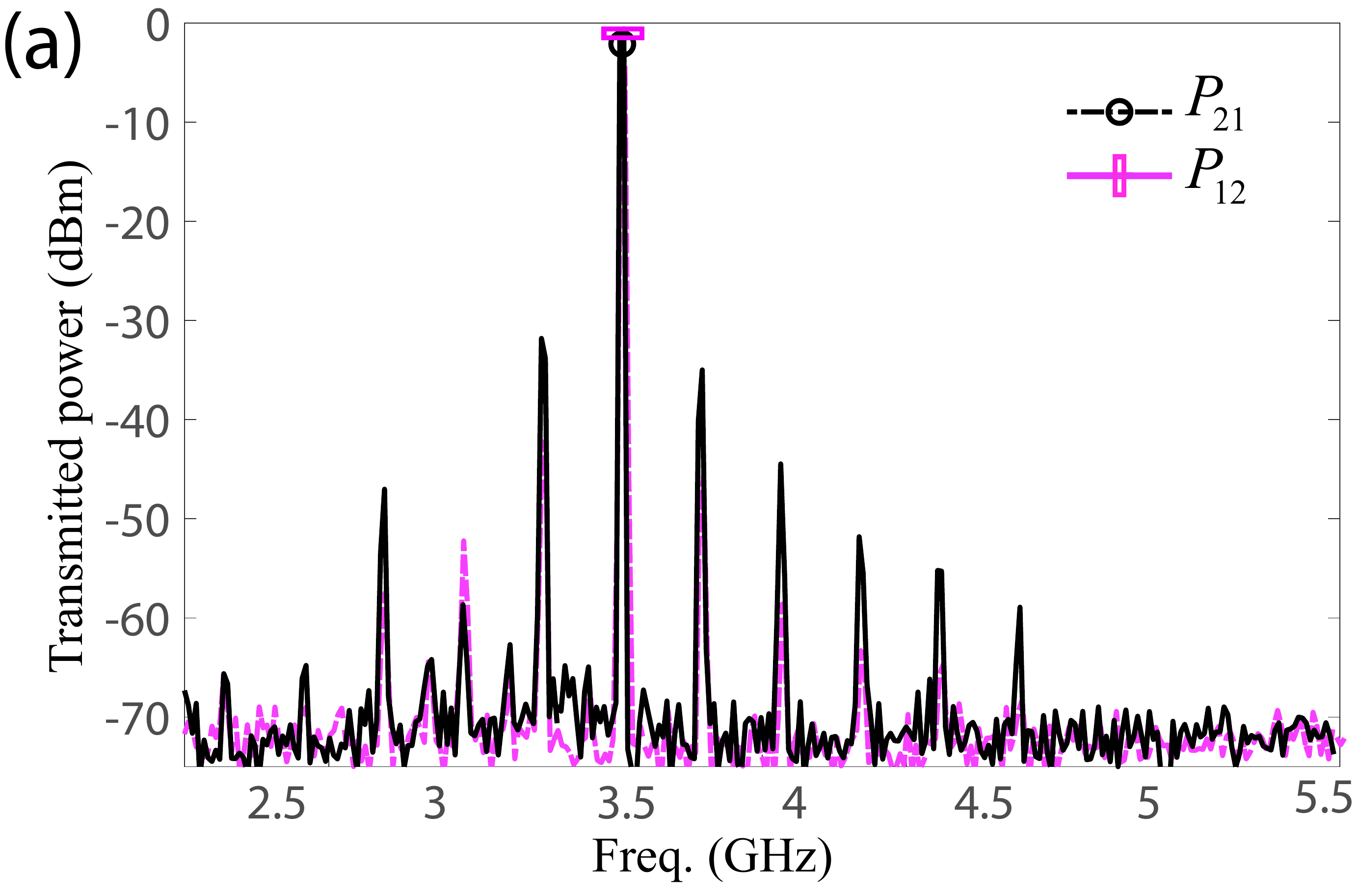}} 
		\subfigure{\label{fig:measb} 
			\includegraphics[width=0.9\columnwidth]{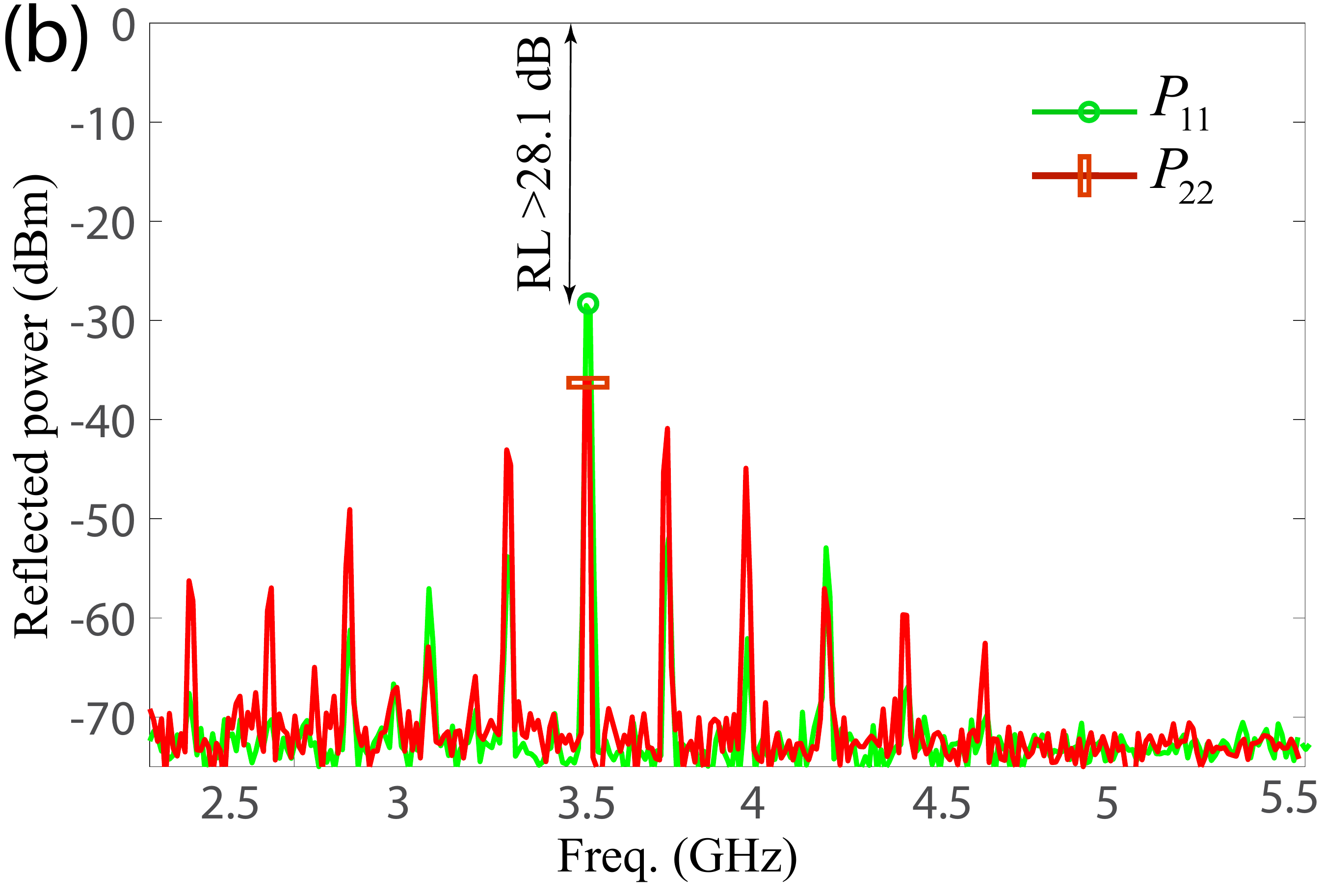}}
		\caption{Experimental results for $\omega_0=3.5$ GHz. (a) Power transmission from port 1 to port 2, $P_{21}$, and from port 2 to port 1, $P_{12}$. (b) Return loss (input matching) at the two ports of the nonreciprocal phase shifter.} 
		\label{Fig:meas}
	\end{center}
\end{figure}

 Figures~\ref{fig:measf} plots the experimental results for wave transmission between the two ports of the nonreciprocal phase shifter. It may be seen from these results that an insertion loss of less than 1.2 dB between the two ports is achieved, whereas the undesired time harmonics are strongly suppressed, i.e., they are 31 dB lower than the main signals. Figure~\ref{fig:measb} plots the return loss at the two ports of the nonreciprocal phase shifter, that is, $P_{11}$ and $P_{22}$, where $P_{11}<-28.1$ dBm ($\text{RL}_{11}>28.1$ dB) and $P_{22}<-34.5$ dBm ($\text{RL}_{22}>34.5$ dB). 

\begin{figure}[h!]
	\centering
	\includegraphics[width=0.9\columnwidth]{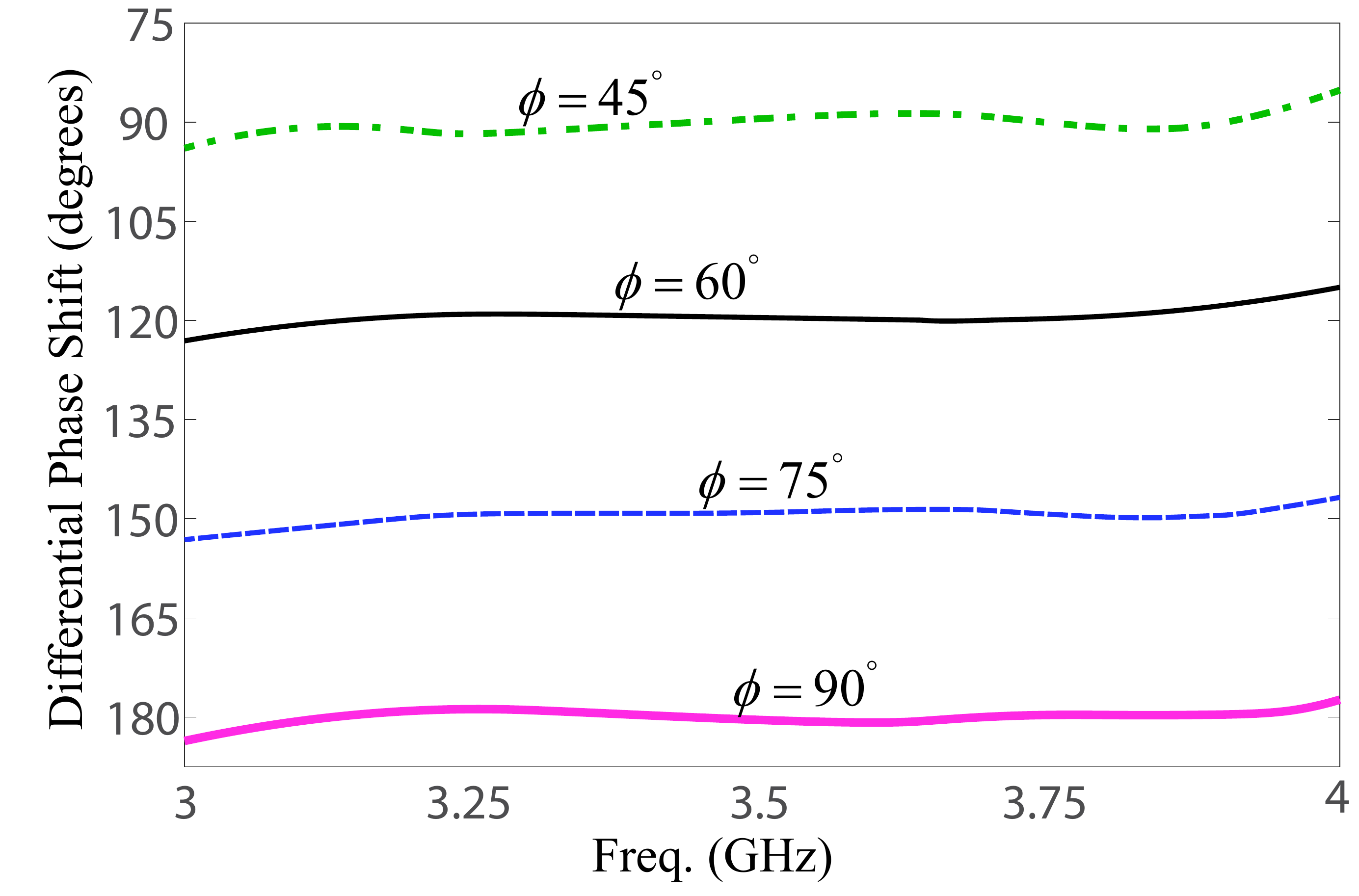}
	\caption{Experimental results for differential phase shift $\Delta \phi=\phi_\text{backward}-\phi_\text{forward}=2\phi$. Here, the input frequency $\omega_0$ is swept from 3 to 4 GHz to show the broad-band response of the temporal nonreciprocal phase shifter.}
	\label{fig:phsh}
\end{figure}

 Figures~\ref{fig:phsh} plots the experimental results for differential phase shift, that is, $\Delta \phi=\phi_\text{backward}-\phi_\text{forward}=2\phi$, where the input frequency $\omega_0$ is swept from 3 to 4 GHz to show the broad-band response of the temporal nonreciprocal phase shifter. This results show that the proposed temporal nonreciprocal phase shift offers a controllable nonreciprocal phase shift in a broad frequency band.

Such a temporal nonreciprocal phase shifter can be built in a more compact fashion, and may be integrated into IC technology for high frequencies thanks to the availability of variable capacitors at high frequencies. In addition, in contrast to transistor-based nonreciprocal phase shifters~\cite{Tanaka_1965} where nonlinear and low power handling transistors are placed in series with the incident signal, our nonreciprocal phase shifter is endowed by high power rating as the varactors are placed in parallel to the incident signal. In our experiment, we applied up to +33 dBm input signal, and the power rating is expected to exceed $+47$~dBm. Having leveraged the linear properties of the time modulation, the proposed time-modulated nonreciprocal phase shifter exhibits an outstanding linear response, where the P1dB is +31.1 dBm.

In conclusion, a magnetless temporal nonreciprocal phase shifter is proposed. The experimental results demonstrate a low insertion loss of less than 1.2 dB which can be further improved. Furthermore, the unwanted time harmonics are 30 dB lower than the main signal. Furthermore, the fabricated nonreciprocal phase shifter exhibits a highly linear response, with the P1dB of +30.4 dBm. In addition, the proposed device can be redesigned for integration into IC technology to achieve much smaller footprint and versatile operation.\\

\textbf{AUTHOR DECLARATIONS}

\textbf{Conflict of Interest}

The authors have no conflicts of interest to declare.\\

\textbf{DATA AVAILABILITY}

The data that support the findings of this study are available
from the corresponding authors upon reasonable request.\\

\textbf{ACKNOWLEDGEMENTS}

This work is supported by the Natural Sciences and Engineering Research Council of Canada (NSERC).

\bibliography{Taravati_Reference}

\end{document}